\documentclass[a4paper,11pt]{article}
\usepackage{pos}

\title{Making cosmic particle accelerators visible and audible}
 \ShortTitle{Making cosmic particle accelerators visible and audible}

\author*[a]{Stefan Ohm}
\author[b]{Konrad Rappaport}
\author[c]{Carsten Nicolai}
\author[a]{Till Mundzeck}
\author[a]{Andrew Taylor}
\author[a]{Sylvia Zhu}
\author[a]{Matthias F\"u{\ss}ling}
\author[d,a]{Robert Daniel Parsons}

\affiliation[a]{DESY, D-15738 Zeuthen, Germany}
\affiliation[b]{Science Communication Lab, Heiligendammer Str. 15, 24106 Kiel, Germany}
\affiliation[c]{Studio Carsten Nicolai, Auguststrasse 26d, 10117 Berlin, Germany}
\affiliation[d]{Institut für Physik, Humboldt-Universit\"at zu Berlin, Newtonstr. 15, D 12489 Berlin, Germany}

\emailAdd{stefan.ohm@desy.de}

\abstract{In a collaboration between astroparticle physicists,
  animation artists from the award-winning Science Communication Lab,
  and musician Carsten Nicolai (a.k.a. Alva Noto), two cosmic particle
  accelerators have been brought to life: the massive binary star Eta
  Carinae, and the exploding star, which resulted in the gamma-ray
  burst GRB190829A. For Eta Carinae, the computer-generated images are
  close to reality because the measured orbital, stellar and wind
  parameters were used for this purpose. Particle acceleration in the
  jet of GRB190829A has also been animated at a level of detail not
  seen before. The internationally acclaimed multimedia artist Carsten
  Nicolai, who uses the pseudonym Alva Noto for his musical works,
  exclusively composed the sound for the animations. The multimedia
  projects aim at making the discoveries more accessible to the
  general public, and to mediate scientific results and their
  reference to reality from an artistic point of view.}

\FullConference{37$^{\rm{th}}$ International Cosmic Ray Conference (ICRC 2021)\\
		July 12th -- 23rd, 2021\\
		Online -- Berlin, Germany}

\begin{document}
\maketitle

\section{Introduction}

Outreach and education are becoming more and more important in
the communciation of scientific results to the public. This effort is
not limited to the communication of interesting discoveries, but
increasingly also to educate the public how science and fundamental
research work in general. In particular astronomy and astrophysics
results attract significant interest. Animations and artist's
impressions of astrophysical objects have long been used to visualise
how objects look that may not be resolvable by telescopes. In recent
years, visualisation and animation software has reached a level of
complexity, which makes it feasible to visualise astrophysical objects
with realistic/physical inputs and to go much beyond the established
``artist impressions''. Although there is no sound in space, the
sonification of an animation would add another dimension to the
audience's experience, making it more immersive. The multimedia
project presented here aims at making the discoveries of two gamma-ray
sources, namely the colliding-wind binary Eta Carinae, and the
gamma-ray burst (GRB) GRB190829A more accessible to the general
public, and to mediate scientific results and their reference to
reality from an artistic point of view. It is an attempt to bring the
areas of science and art together. The paper is organised as
following: some background of how the multimedia project was realised
is given in section \ref{sec:process}. The multimedia animation of Eta
Carinae is presented in Chapter~\ref{sec:etacar}. The physics inputs,
level of realism as well as the animation visualisation and
sonification are discussed in more detail. Chapter~\ref{sec:grb} will
present the multimedia animation of GRB190829A in the same manner and
also highlight differences to the Eta Carinae project in terms of the
practical realisation. Chapter~\ref{sec:outlook} will discuss the
lessons learnt from the cooperation and outline future prospects of
projects like this.

\section{The process of developing and realising the multimedia
  project}
\label{sec:process}
DESY has established a cooperation with the design studio Science
Communication Lab~\citep{scicomlab} in Kiel a few years ago. High-impact scientific
results and other projects are communicated to the public through
innovative visual animations. For the two projects presented in this
work, the cooperation was extended to include Carsten Nicolai
a.k.a. Alva Noto~\citep{alvanoto} to compose soundtracks for the
animations. The baseline for the animations is usually the scientific
publications, which introduce the science questions that have been
studied and present the scientific results. From this input, a first
script for the animation is developed, which introduces the audience
to the astrophysical object, the measurement and the detection with 
instruments on Earth. Animation artists from Science Communication Lab
and scientists involved in the measurement, together develop a script
for the animation. At this point the main source of information needs
to be chosen. There are two different inputs that we used for the Eta
Carinae and GRB190829A animations. For Eta Carinae, mainly real data
was used to capture the dynamics of the system (see
Chapter~\ref{sec:etacar}), whereas for GRB190829A, physical processes
as implemented in the animation software have been used to approximate
conditions in the GRB. The software used to create the animation is
\textit{MAXON Cinema4D}~\citep{maxon} for 3D computer animation,
modeling, simulation, and rendering. The
\textit{xParticles}~\citep{xparticles} plugin was used to support
Newtonian Gravity as well as fluid dynamics (e.g. fire, smoke, or 
grains). Additionally, also the
\textit{TurbulenceFD}~\citep{turbulencefd} fluid dynamics plugin has
been used. Individual scenes are then produced in a wireframe model,
iterated among the participants and at the end put together to form
the full animation. Once the full animation including all scenes was
available as wireframe model, Carsten Nicolai started to compose the
soundtracks. Some final adjustments to the animation and sound were
done before the full rendering of the video was initiated. The two
multimedia projects are discussed in more detail in the following.

\section{Visualisation and sonification of Eta Carinae's gamma-ray
  emission}
\label{sec:etacar}
\begin{figure}[ht]
  \centering
    \includegraphics[width=0.5\textwidth]{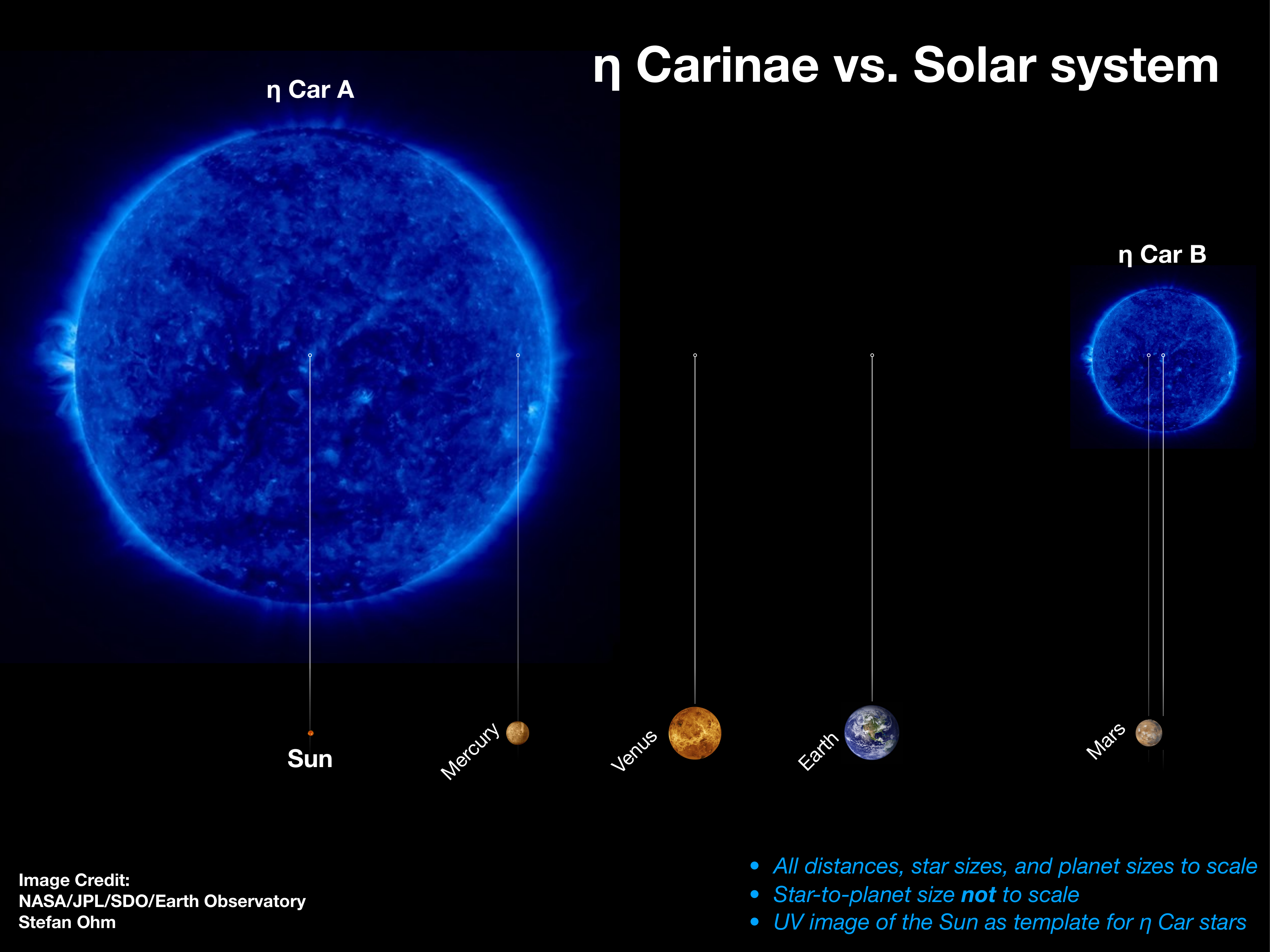}
    \includegraphics[width=0.4\textwidth]{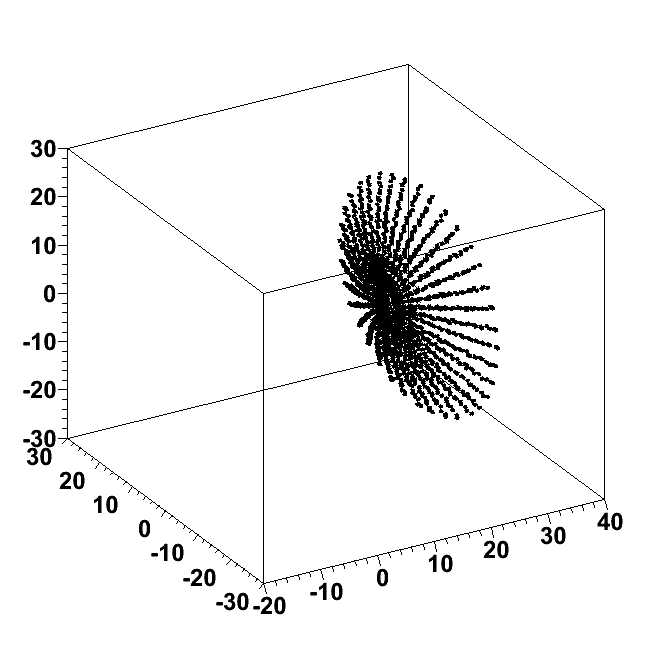}
    \caption{Left: Sketch of the Eta Carinae binary system during periastron
    compared to the solar system. Right: Grid points used to map the
    time-dependent shape of the shock cap, based
    on~\citep{etacar:canto1996}.}
    \label{fig:etacar_input}
\end{figure}

Eta Carinae is a massive stellar binary system 7500 lightyears away
from Earth, and situated in the Carina arm of the Milky Way. It is one
of the most enigmatic objects in the night sky and has been studied as
early as in 1595 when a Dutch expedition to the East Indies with
navigator Pieter Keyser mapped the southern stars
\citep{etacar:dekker1987}. Nowadays, astronomical measurements across all
wavelengths suggest that the system is composed of two very massive
stars ($\sim 100$ and $\sim$20 times the mass of the Sun) that orbit
each other on a very eccentric orbit every 5.54 years. During the
closest encounter, the stars approach each other at 
a distance corresponding to the distance between the Sun and
Mars, (Fig.~\ref{fig:etacar_input}, left). The conditions in this
system are very extreme with the more massive star losing almost an
entire solar mass of material within only 2000 years due to its dense
winds. When the two stellar winds collide they form a colliding wind
region in which gas is heated to temperatures, where the gas shines in
X-rays. At the same time wind particles are accelerated to
Teraelectronvolt energies in shocks in the colliding wind region and
produce gamma rays. Gamma rays have been detected with the LAT
instrument onboard of the Fermi satellite
\citep{etacar:fermipaper2020} and, more recently, with the
H.E.S.S. telescopes situated in Namibia
\citep{etacar:hesspaper2020}. It is the first system of this kind
detected at such high energies and the origin of the gamma-ray
emission remains somewhat elusive.

\subsection{Realism and physics input}

The aim of the visualisation was to strike a balance between reaching
a certain level of realism while being visually appealing and
not overloading the viewer with information. The most important
ingredients to capture the dynamics of the stars and their extreme
properties were provided by using the measured orbital parameters and
inferred stellar properties \citep[see][and references
therein]{etacar:ohm2015}. Thereby we were able to show the
time-dependent movement of the stars along their orbits and with the
realistic speeds (the stars move faster when they are closer). The
second major ingredient was the colliding wind region, which was
derived using the Canto et al. (1996) \citep{etacar:canto1996}
representation, who make use of the momentum balance between the two
stellar winds. The shock cap was parametrised using a grid of 36
azimuthal and 82 radial points and used to guide the
simulation (Fig.~\ref{fig:etacar_input}, right). Together with the
stars, the shock cap is not static, but moving in space, and the gas
inflowing from the two stars mixes and is ejected outwards
ballistically from the edge of the shock cap. A limitation of the
animation was reached when showing the gas density difference between
the two stellar winds and the much denser colliding wind region. Here
a compromise had to be made to maintain the undisguised view into the
entire system and all its components. With realistic density
contrasts, one or multiple system components would have been hidden.

\subsection{Output and media coverage}

\begin{figure}[ht]
    \centering
    \includegraphics[width=0.45\textwidth]{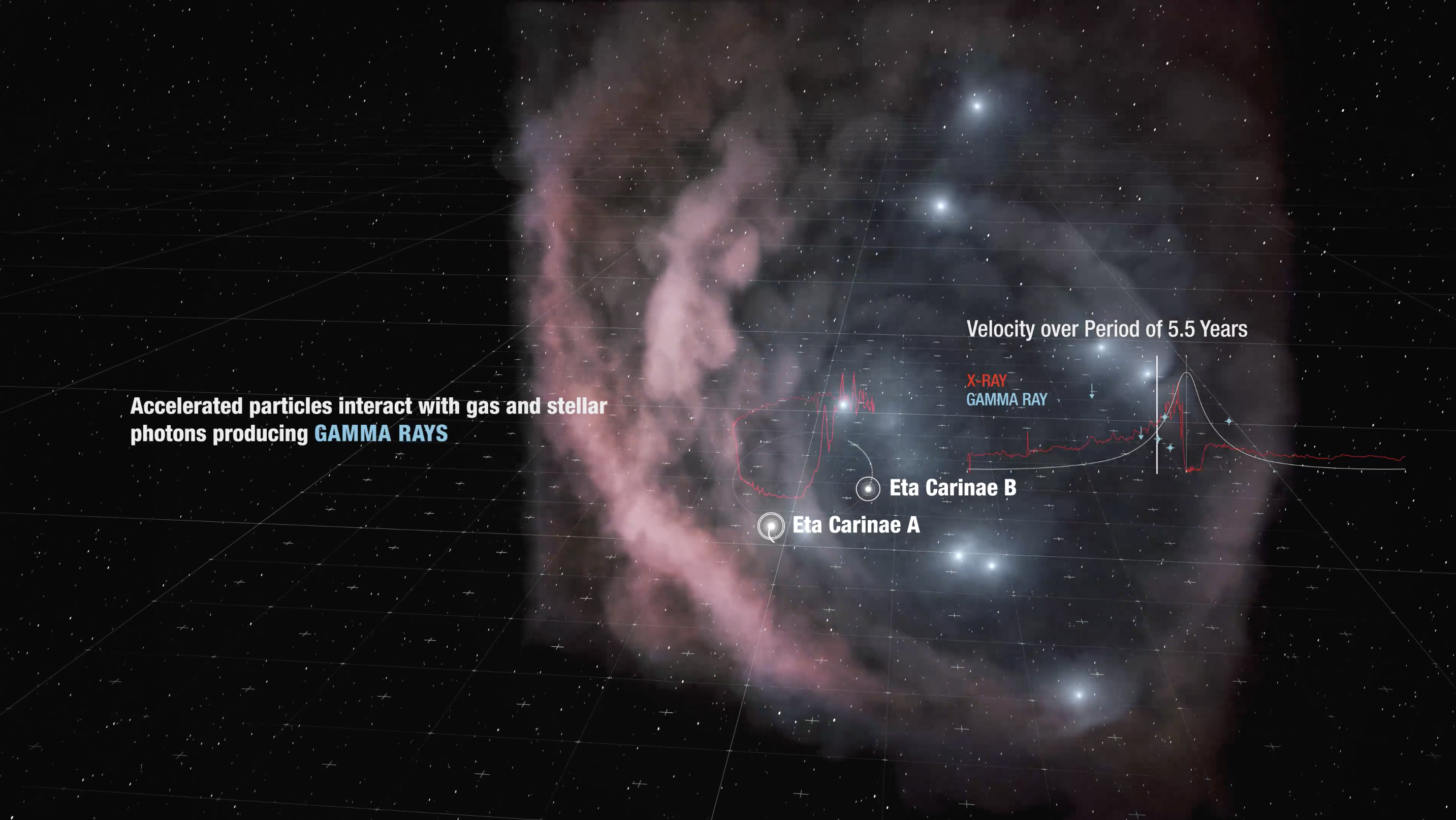}
    \includegraphics[width=0.45\textwidth]{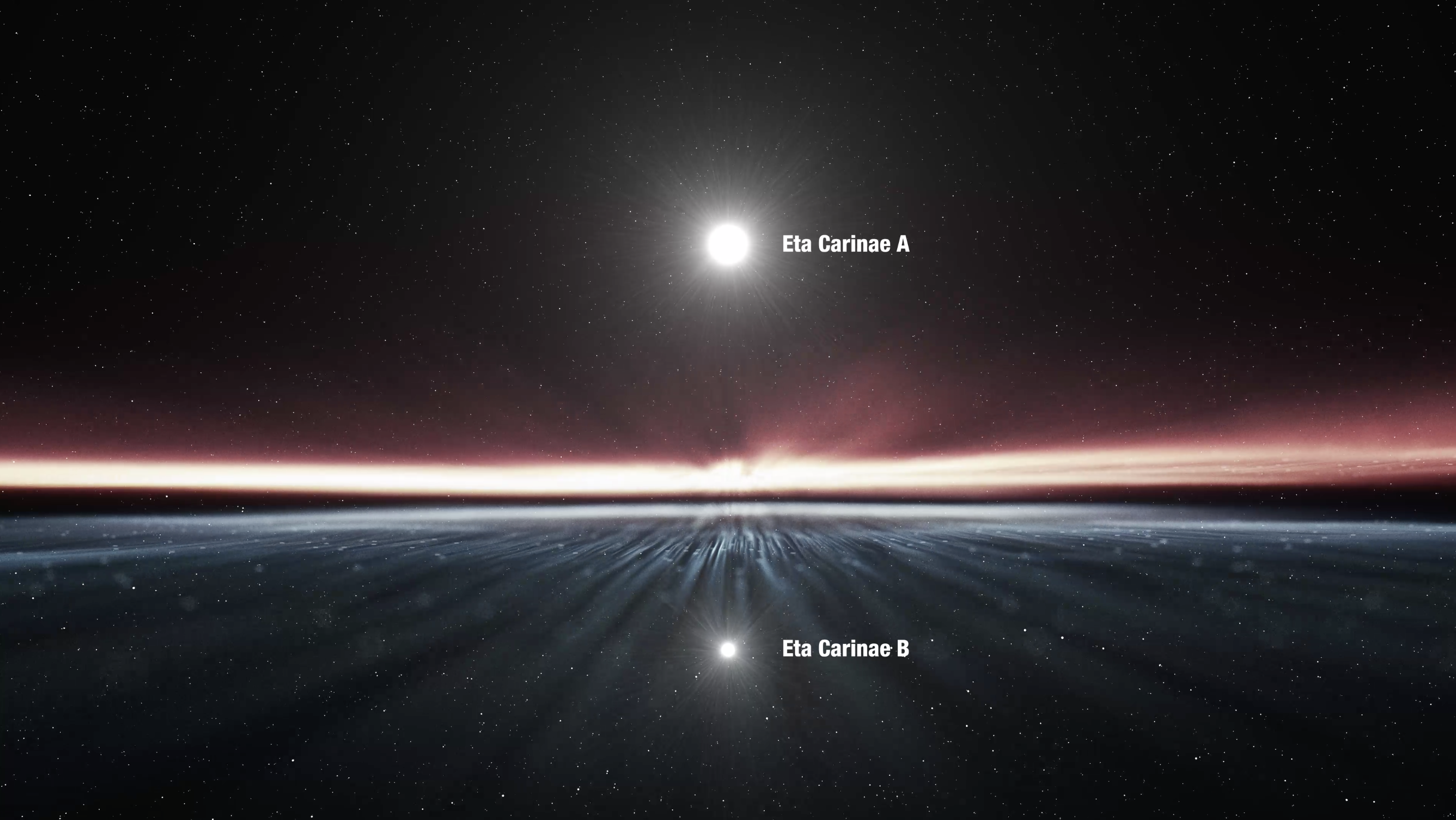}
    \caption{Two extracted stills from the Eta Carinae animation,
      showing the top-down view into the system, including the stars,
      colliding wind region, gamma-ray emission as well as explanatory
      text (left). A side-view into the colliding wind region is shown
      with the two star sizes to scale (right). Copyright: DESY,
      Science Communication Lab.}
    \label{fig:etacar_still}
\end{figure}

The final animation comprises 3:40 minutes of runtime, includes
embedded text in two languages (English \citep{etacar:ytvideo2020e}
and German \citep{etacar:ytvideo2020d}), and a
picture-in-picture view of the X-ray and gamma-ray
lightcurve along the orbit. The different parts of the system and the
scientific results are built up one after another, starting with the
two stars, their orbit, their winds, the colliding wind region, and
the production of X-ray and gamma-ray emission. Apart from a
top-side view into the system (Fig.~\ref{fig:etacar_still}, left),
where the focus point is towards the center-of-mass of Eta Carinae,
one scene towards the end is set in the shock cap, with the focus
point at the contact discontinuity. Material from both stellar winds
flows towards the viewer on the top/bottom side of the screen
(Fig.~\ref{fig:etacar_still}, right). The gamma-ray emission is
visualised through individually produced photons that originate partly
in the colliding wind region, and partly in the outflowing
material. Thereby, we tried to capture the still uncertain origin of
the gamma-ray emission. Given the complexity of the content that is
conveyed to the viewer, explanatory text was embedded in
the video itself and meant to guide the viewer through the different
parts of the animation. To anker the view, a 3D grid of coordinates
was put in the animation to provide a reference frame. 

The sonification was based on the final animation, which pre-defined
the appearance and disapperance of different sound elements as well
as their transition to form the final soundtrack. After an initial
exchange on the content of the animation (the stars, the winds,
particle acceleration in magnetic fields, and the extreme conditions
in the system) Carsten Nicolai developed the soundtrack from a sound
artist's perspective. The process of developing the soundtrack and
Carsten Nicolai's view have been captured in a interview that was
released together with the press release and animation
\citep{etacar:alvanotointerview2020}. The English and German YouTube
videos have been viewed $\sim$13400 and $\sim$2800 times,
respectively. 

\section{Visualisation and sonification of the gamma-ray burst
  GRB190829A}
\label{sec:grb}

Triggered by the detection of X-ray emission from a long GRB using the
Swift satellite, the H.E.S.S. telescopes initiated follow-up
observations from the ground in the gamma-ray regime as soon as
GRB190829A was visible to H.E.S.S. under favourable conditions. A
real-time analysis of the data taken on site resulted in a detection of a new
gamma-ray source at the position of this GRB. The gamma-ray signal was
detected also in the following two nights, with a fading
intensity. Spectroscopic observations established GRB190829A to be
much closer to Earth than typical GRBs. Moreover, the GRB emission was
later accompanied by an afterglow from the subsequent supernova
explosion.  The temporal and spectral X-ray and gamma-ray measurements
challenge current models for the non-thermal emission from long GRBs
and have been published in the \textit{Science} journal
(see~\citep{grb190829a:hesspaper2021} for further details and
  references).

\subsection{Realism and physics input}

This unique discovery and its physics implications motivated the
second cooperation between Science Communication Lab, Carsten Nicolai
and DESY members. Similar to Eta Carinae, the aim was to bring the
scientific result to life, make it more accessible to the general
public, and to mediate its reference to reality from an artistic
point of view \citep{grb190829a:desypr2021}.

\begin{figure}[ht]
    \centering
    \includegraphics[width=0.469\textwidth]{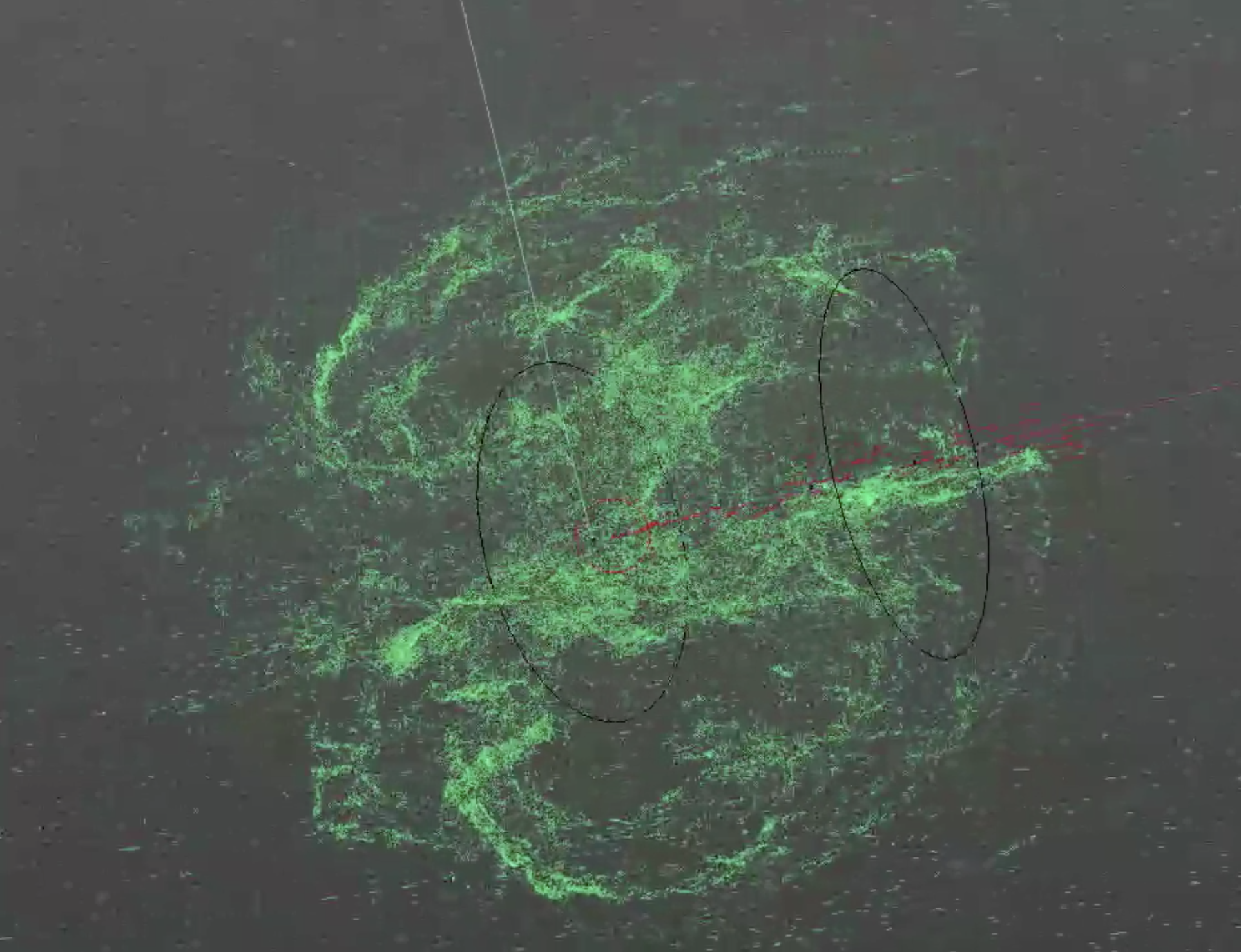}
    \includegraphics[width=0.431\textwidth]{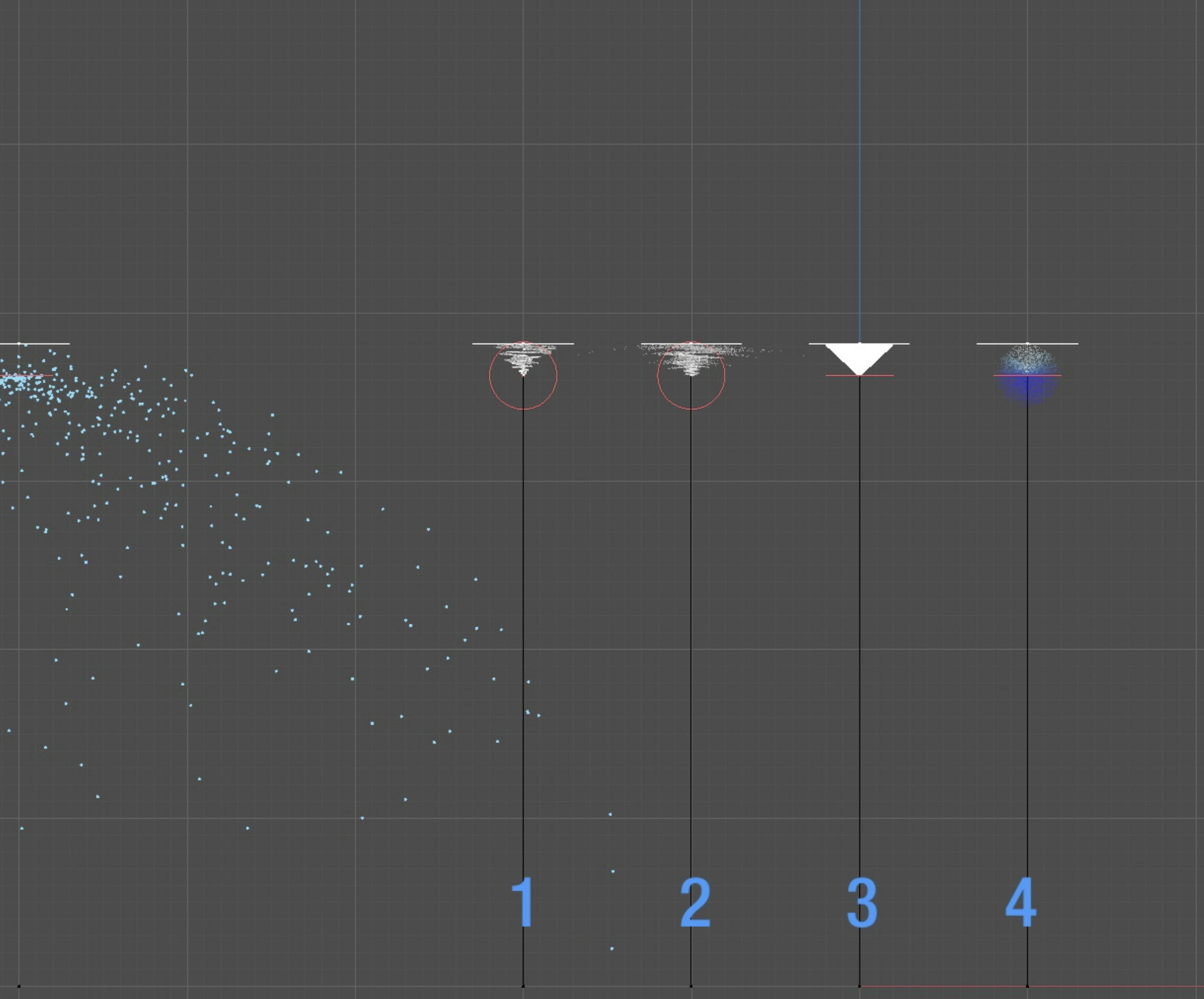}
    \caption{Two wireframe screenshots of individual scenes in the
      animation. The left hand side shows the gas cloud to which a
      test particles is coupled and gets deflected. The right-hand side
      shows different attemps at illustrating the relativistic beaming
      and emission of photons from the head of the jet. Copyright:
      DESY, Science Communication Lab.}
    \label{fig:grb_input}
\end{figure}

Due to the cosmological distance to GRB190829A, no telescope could
resolve the system, which is why the visualisation was mainly guided
by unresolved (but time-dependent) measurements from optical, X-rays and
gamma rays, as well as by established theoretical models. The size of
the collapsing star with respect to the jet and the jet opening angle
are inspired by measurements and in particular the
relativistic boosting factor, $\Gamma$. The shape of the jet are
guided by theoretical models of jet breakouts like in
\citep{grb:zhang2004}. The head of the jet depicts the different
regions where particles are believed to be accelerated, i.e. the
forward and reverse shocks; and the contact discontinuity. The
outflowing material in the jet was realised through a fluid
simulation, which is constantly fed with new material. This
representation does not necessarily resemble the likely situation in 
the GRB, as it is believed that energy was fed into the jet in the
launching event and that particles are picked up from the interstellar
medium and subsequently get accelerated. Instead, we use a test particle
that is coupled to the outflowing jet material, is deflected in
magnetic fields and emits synchrotron
emission. Figure~\ref{fig:grb_input} (left) shows a wireframe snapshot
of the outflowing gas cloud, with a test particle emitting
photons. The emission of synchrotron photons predominatly in the
forward direction reflects the relativistic beaming
effect. Figure~\ref{fig:grb_input} (right) shows different 
realisations of the relavistic beaming in the emitted photons. The
rapid radial drop of intensity of the gamma rays that are launched
from the head of the jet follows current state-of-the-art models as
well. 

\begin{figure}[ht]
    \centering
    \includegraphics[width=0.33\textwidth]{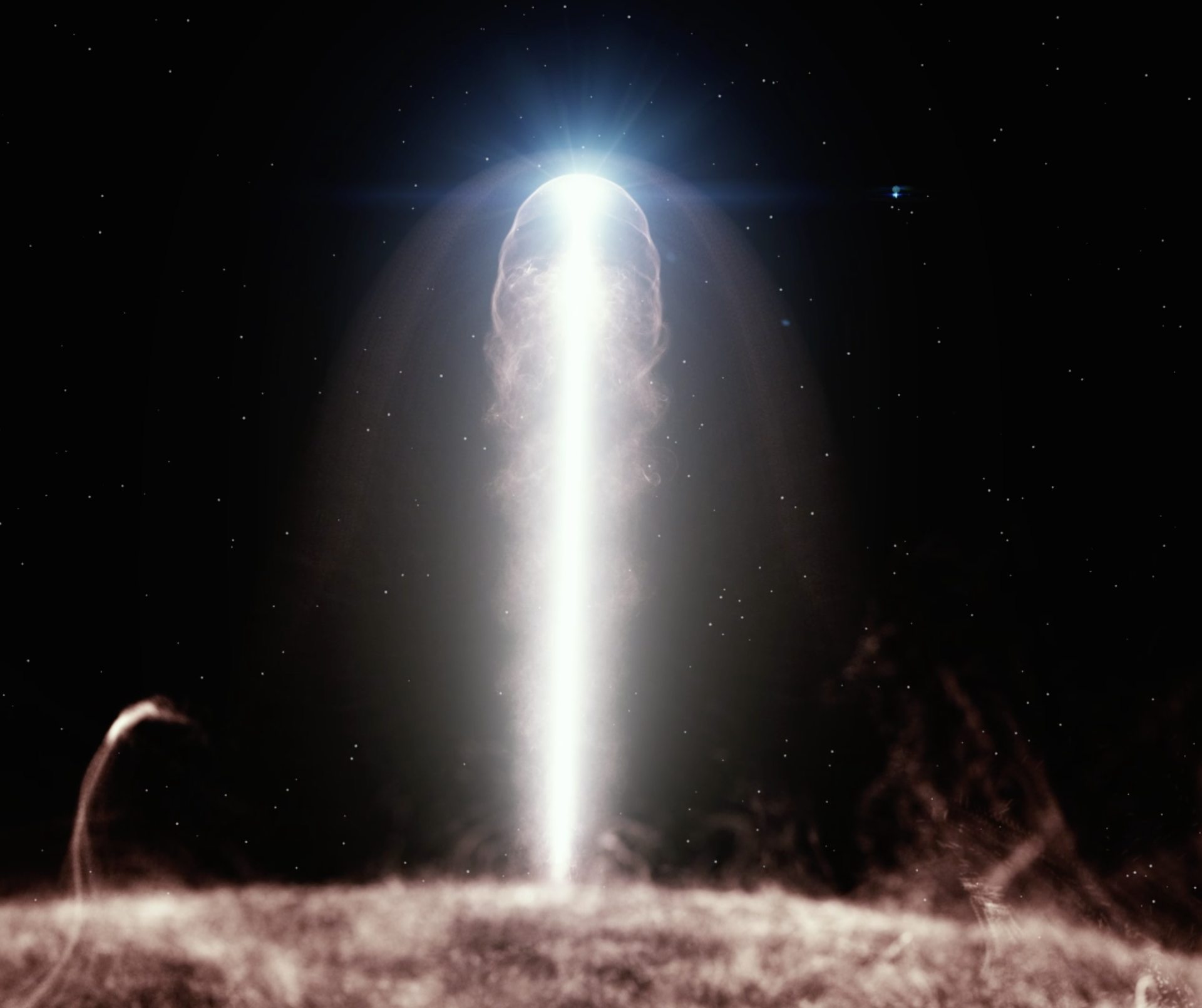}
    \includegraphics[width=0.3\textwidth]{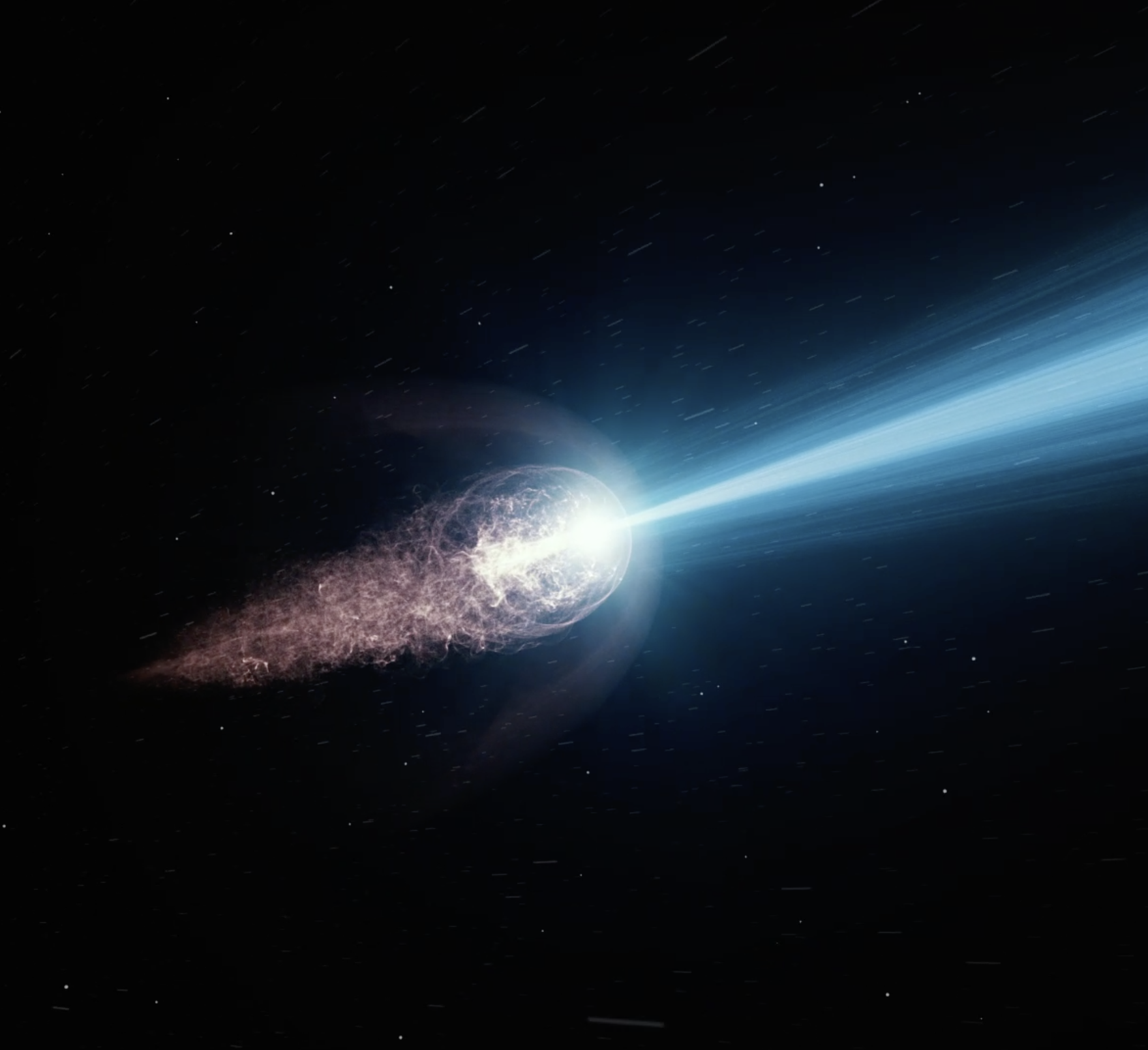}
    \includegraphics[width=0.33\textwidth]{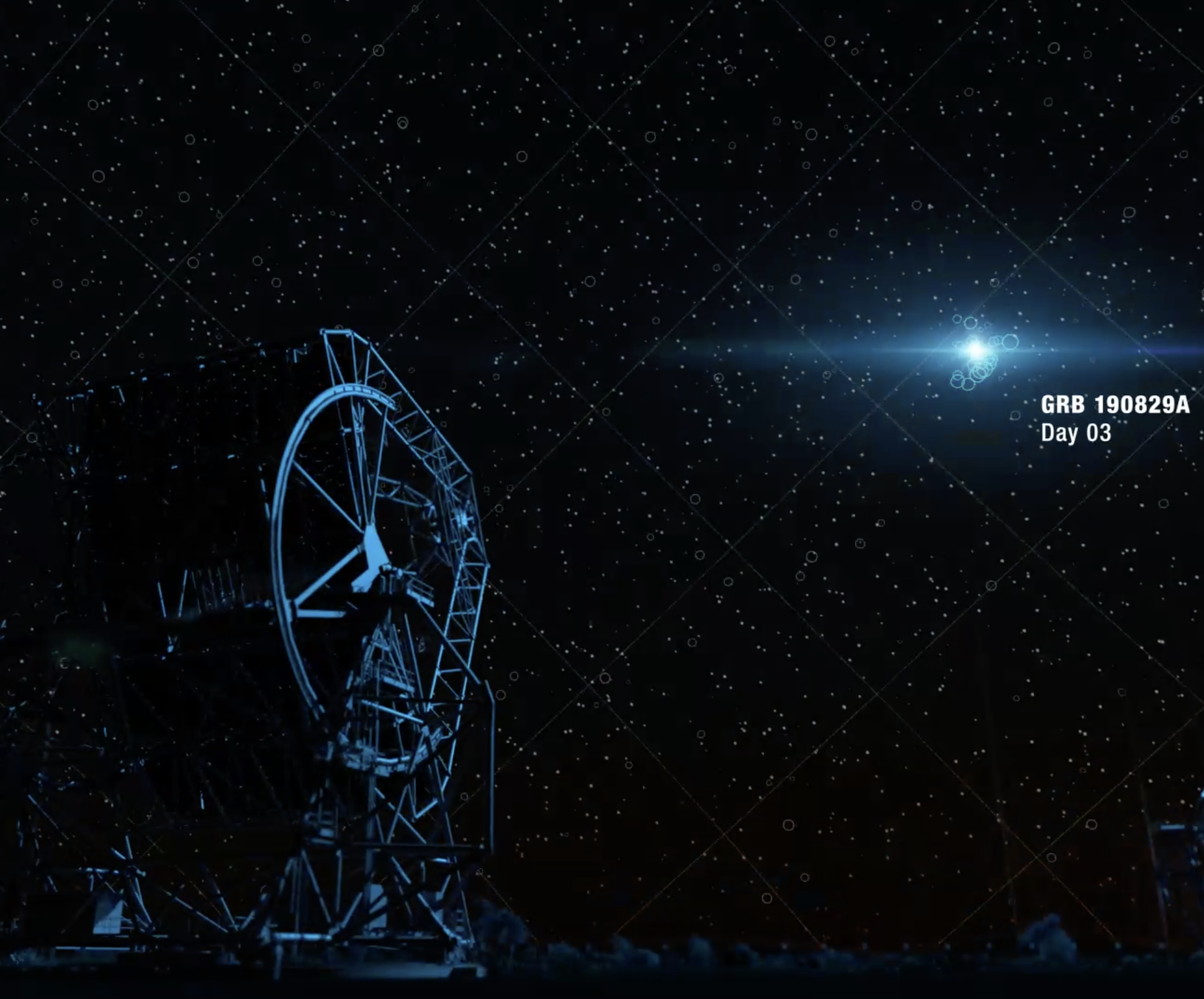}
    \caption{Shown are three screenshots of the jet breakout from the
      star and its launch (left), the propagation of the jet, particle
      acceleration and gamma-ray production (middle), and the
      detection with the H.E.S.S. telescopes (right). Copyright: DESY,
    Science Communication Lab.}
    \label{fig:grb_output}
\end{figure}

The visualisation of the X-ray and gamma-ray photons arriving on Earth
adds another layer of realism. CAD models of the Swift
satellite and the H.E.S.S. telescopes give the viewer a much more
realistic understanding of how the instruments scientists operate look
like. The space-view towards Earth features the Swift satellite in the
foreground, which detects X-rays, but also shows a realistic
geographical map of Namibia and South Africa in the
background. Very-high-energy photons are punching into Earth's
atmosphere and initiate particle showers, which shine in blue
light. The size of the particle showers are not to scale as they would
not be visible from such a large distance. Finally, the movement of
the H.E.S.S. telescopes tracking the GRB over three nights, while being
parked in during the day, is also a realistic resemblence of the
actual observations and how they have been conducted. The three main
elements of the animation, namely the jet breakout and launch, the jet
propagation, particle acceleration and gamma-ray production, as well
as the detection by instruments on Earth are displayed in three
screenshots in Figure~\ref{fig:grb_output}.

\subsection{Output and media coverage}

The final animation has a length of 1:50 minutes. Shorter
versions for distribution on social media have been produced --
most importantly two reels of 30-second length that were used on
Instagram. As for the Eta Carinae video, the animation has been
uploaded to YouTube, but in this case using the subtitle option of
YouTube itself, rather than embedding the text in the video. This has
the advantage that one can use one video link for multi-lingual
distribution and to support visually impaired viewers. The YouTube
video \citep{grb:ytvideo2021a}, which was released at the same time as
the \textit{Science} paper and accompanied by extensive press coverage 
(e.g. \citep{grb190829a:desypr2021}) was viewed more than 220,000
times in less than two weeks. On average, people watched 80\% of the
video, indicating that those that didn't stop the playback
immediately, most likely watched the animation till the end.

The sonification by Carsten Nicolai was again based on the final
animation, which pre-defined the different sound elements and their
transition to form the final soundtrack. Also for this project,
Carsten Nicolai developed the soundtrack from a sound artist's
perspective. For the GRB190829A soundtrack, an underlying beat was
included that should take the viewer/listener through this highly
dynamic system. A dedicated social media campaign on Instagram and
Twitter highlighting the sound aspect of the cooperation was developed
and realised in cooperation between the social media teams of Carsten
Nicolai, DESY, and Science Communication Lab. The animation including
the soundtrack was uploaded as a second YouTube video on July 1st,
2021~\citep{grb:ytvideo2021b} resulting in more than 1100 views in four
weeks. The Instagram feature received more than 3000 views via the
Alva Noto, Science Communication Lab and DESY channels.

\section{Future prospects}
\label{sec:outlook}

To our knowledge, this is the first multimedia project of this kind in
astrophysics. It offers a unique way to communicate scientific results
beyond the astronomy-interested public, and aims at approaching
an audience with a specific interest in (audio-visual) art. The
cooperation between Science Communication Lab, Carsten Nicolai and
DESY scientists is planned to be continued and feature other cosmic
particle accelerators. In the future we plan to approach projects like
this even more from the sound perspective, by e.g. composing the
soundtrack first and compiling the visualisation (and scientific
content) based on the soundtrack.

%
%
%


\begin{thebibliography}{99}

\bibitem{scicomlab}
  Science Communciation Lab,
  Design Studio, 
  accessed 05.07.2021,
  \url{https://www.scicom-lab.com}
  
\bibitem{alvanoto}
  Carsten Nicolai / Alva Noto,
  Studio Carsten Nicolai, 
  accessed 27.07.2021,
  \url{https://www.alvanoto.com}
  
\bibitem{maxon}
  Maxon Cinema 4D,
  3D computer animation, modeling, simulation, and rendering software, 
  accessed 05.07.2021,
  \url{https://www.maxon.net/en/cinema-4d}

\bibitem{xparticles}
  INSYDIUM,
  X-Particles plugin for Maxon: Cinema 4D, 
  accessed 05.07.2021,
  \url{https://insydium.ltd/products/x-particles}

  
\bibitem{turbulencefd}
  JAWSET,
  TurbulenceFD plugin for Maxon: Cinema 4D, 
  accessed 05.07.2021,
  \url{https://www.jawset.com/turbulencefd}
  
\bibitem{etacar:dekker1987}
Dekker, E., Early explorations of the southern celestial
sky. \textit{Annals of Science}, 44: 439–470 (1987)

\bibitem{etacar:fermipaper2020} 
White, R.; Breuhaus, M.; Konno, R., Ohm, S., Reville, B., and Hinton,
J.A., Gamma-ray and X-ray constraints on non-thermal processes in
$\eta$~Carinae, \textit{A\&A} 635A, 144W (2020).

\bibitem{etacar:hesspaper2020} 
H.E.S.S. Collaboration, Abdalla, H.; Adam, R., et al., Detection of
very-high-energy $\gamma$-ray emission from the colliding wind binary
$\eta$~Car with H.E.S.S., 
\textit{A\&A} 635A, 167H (2020).

\bibitem{etacar:ohm2015}
  Ohm, S., Zabalza, V., Hinton, J.A. \& Parkin, E.R., On the origin of
  $\gamma$-ray emission in $\eta$ Carina, \textit{MNRAS Letters}, 449 (1), L132 –
  L136 (2015)

\bibitem{etacar:canto1996}
Canto J., Raga A. C., \& Wilkin F. P., Exact, Algebraic Solutions of
the Thin-Shell Two-Wind Interaction Problem, \textit{ApJ}, 469, 729 (1996)

\bibitem{etacar:ytvideo2020e}
  YouTube,
  Eta Carinae - a new source of very high-energy cosmic gamma
  radiation,
  accessed 27.07.2021,
  \url{https://www.youtube.com/watch?v=uUFJXjIhUkQ}

\bibitem{etacar:ytvideo2020d}
  YouTube,
  Eta Carinae - eine neue Quelle sehr energiereicher kosmischer
  Gammastrahlung, 
  accessed 27.07.2021,
  \url{https://www.youtube.com/watch?v=IWOxekXsypU}
  
\bibitem{etacar:alvanotointerview2020}
  DESY News,
  ``When I see gamma rays, they automatically have a sound for me'',
  accessed 30.06.2021,
  \url{https://www.desy.de/news/news_search/index_eng.html?openDirectAnchor=1865}

\bibitem{grb190829a:hesspaper2021} 
H.E.S.S. Collaboration, Abdalla, H.; Aharonian, F.A, Ait Benkhali, F.,
et al., Revealing x-ray and gamma ray temporal and spectral similarities
in the GRB 190829A afterglow, \textit{Science} 372, 6546, 1081--1085
(2021).

\bibitem{grb190829a:desypr2021}
  DESY News,
  Binary star as a cosmic particle accelerator,
  accessed 30.06.2021,
  \url{https://www.desy.de/news/news_search/index_eng.html?openDirectAnchor=2080}

\bibitem{grb:zhang2004}
  Zhang, W. \& Woosley, S. E., The propagation and eruption of
  relativistic jets from the stellar progenitors of GRBs,
  \textit{ApJ}, 608, 365 (2004)

\bibitem{grb:ytvideo2021a}
  YouTube,
  Gamma-ray Burst in Our Cosmic Backyard / Gammablitz aus der
  kosmischen Nachbarschaft,
  accessed 30.06.2021,
  \url{https://www.youtube.com/watch?v=VdF4gknH9YM}

\bibitem{grb:ytvideo2021b}
  YouTube,
  A cosmic experience - Gamma-ray Burst GRB 190829A,
  accessed 27.07.2021,
  \url{https://www.youtube.com/watch?v=oHEpiv33fZQ}


\end{thebibliography}
\end{document}